\documentclass[preprint,12pt]{elsarticle}

\usepackage{amsmath,amssymb,amsfonts}
\usepackage{graphicx}
\usepackage{textcomp}
\usepackage{xcolor}
\usepackage{microtype}
\usepackage{algorithm}
\usepackage{algpseudocode}
\usepackage{booktabs}

\journal{Digital Signal Processing}

\begin{document}

\begin{frontmatter}

\title{Fourier Analysis of Signals on Directed Acyclic Graphs (DAG) Using Graph Zero-Padding}    

\author[ucg]{Ljubi\v{s}a Stankovi\'{c}\corref{cor1}} 
\ead{ljubisa@ucg.ac.me}
\author[ucg]{Milo\v{s} Dakovi\'{c}} 
\ead{milos@ucg.ac.me}
\author[ucg,pgu]{Ali Bagheri Bardi} 
\ead{alibagheri@ucg.ac.me, bagheri@pgu.ac.ir}
\author[ucg]{Milo\v{s} Brajovi\'{c}} 
\ead{milosb@ucg.ac.me}
\author[ucg]{Isidora Stankovi\'{c}} 
\ead{isidoras@ucg.ac.me}
\cortext[cor1]{Corresponding author.}
\affiliation[ucg]{organization={University of Montenegro},
            city={Podgorica},
            postcode={81000}, 
            country={Montenegro}}
\affiliation[pgu]{organization={Persian Gulf University},
            city={Bushehr},
            postcode={7516913817}, 
            country={Iran}}

\begin{abstract}
	Directed acyclic graphs (DAGs) are used for modeling causal relationships, dependencies, and flows in various systems. However, spectral analysis becomes impractical in this setting because the eigendecomposition of the adjacency matrix yields all eigenvalues equal to zero. This inherent property of DAGs results in an inability to differentiate between frequency components of signals on such graphs. This problem can be addressed by {alternating the Fourier basis or adding edges in a DAG}. However, these approaches change the physics of the considered problem. To address this limitation, we propose a \textit{graph zero-padding} approach. This approach involves augmenting the original DAG with additional vertices that are connected to the existing structure. The added vertices are characterized by signal values set to zero.
	The proposed technique enables the spectral evaluation of system outputs on DAGs (in almost all cases), that is  the computation of vertex-domain convolution without the adverse effects of aliasing due to changes in a graph structure, { with the ultimate goal of preserving the output of the system on a graph as if the changes in the graph structure were not performed}.
\end{abstract}


\end{frontmatter}

\section{Introduction}
  Graph signal processing is an emerging research area at the intersection between signal processing and graph theory, dealing with the analysis, processing, and interpretation of data defined on graphs \cite{Sig_Graphs2023, Graph_filters, DisctSig_Graphs, Graphs_1, trends1, trends2, trends3, LS_Graph}. In many real-world applications, either the data domain or the data structure can be represented by graphs, with edges representing relationships or interactions among data points \cite{trends2,LS_Graph}. 
  
  One particular type of graph that is of great significance in various areas is the Directed Acyclic Graph (DAG) \cite{CausalFourier, FastMobius, Mobius, SparseDAG, DAG-GCN, ChenS}. The DAG is a special type of a directed graph, whose main characteristic is that it is free of cycles. DAGs are commonly used to model causal relationships, dependencies, and flows in various systems \cite{CausalFourier}. In this context, DAGs are used to model and analyze dynamic systems where information (or signal) flows from one vertex to another, causing a cascade of effects. This is particularly relevant in the fields such as epidemiology, finance, neuroscience, and machine learning, where understanding  causal relationships between variables or events can lead to better predictions and interventions \cite{CausalFourier,DAG-GCN,GCNN}. 
  
  Spectral analysis and processing based on the standard Graph Fourier Transform (GFT) are challenging for signals on DAGs, since all eigenvalues of the adjacency matrix are equal to zero, rendering the frequency components of the analyzed signal indistinguishable \cite{CausalFourier}. 
  The majority of endeavors aimed at introducing an appropriate candidate for the GFT concentrate on altering the Fourier basis for Jordan Normal Form (JNF) computation, which can significantly influence the graph signal processing methods \cite{21,22,23,24,25,26,27,28,29}. To tackle the spectral analysis issue, while preserving the integrity of the Fourier basis,  an algorithm has recently been proposed in   \cite{seifert2021digraph}. The core concept entails strategically adding optimal edges to the graph until all non-trivial Jordan blocks are eliminated. This approach inherently changes the physics of the graph signal processing problem and the output of a system on this graph. It is also important to emphasize that, in practical scenarios, the identification of suitable edges is contingent upon having access to the Jordan normal form of the adjacency matrix.
  	
  	 The main contribution of this study is  introducing an algorithm to address the gap identified in \cite{seifert2021digraph}. This is achieved through implementing two operations. Firstly, the method of adding edges is altered. Instead of computing the JNF, the topology of the DAG is considered. Source vertices are identified in each iteration, and through appropriate connections between them, a graph with only one sink and source is formed, directing us to a Hamiltonian cycle. The introduction of new edges leads to a change in topology. By adding some appropriate vertices, and leveraging the Hamiltonian cycle, the same output of a system on graph signals on the original graph and the modified shapes can be preserved.

   To this aim, we introduce a novel concept called \textit{graph zero-padding}, inspired by the zero-padding in classical signal processing \cite[p. 591]{DSPbook1}, \cite{DSPbook}. Graph zero-padding is implemented  by adding some new, appropriate,  vertices connected to the existing graph structure, with the signal values on those vertices set to zero.  This concept opens the possibility to calculate the output of a graph system (filtering based on convolution), without aliasing effects (keeping the original DAG output as if the changes in the graph structure were not done).    The proposed concept, along with the idea of forming a Hamiltonian path in a general DAG, is used to define a simple algorithm  that can simultaneously eliminate the Jordan block associated with the eigenvalue zero.  Indeed, through the implementation of this algorithm, all eigenvalues of the adjacency matrix of the zero-padded graph are non-zero. In the vast majority of situations, the eigenvalues will be distinct, indicating the achievement of diagonalizability, with only a few rare exceptions, that require further attention.
    Diagonalizability of the adjacency matrix allows the GFT analysis and application of various spectral tools on the considered graphs \cite{VF}.
  A simple interpretation of the speed of change (discrete frequency) in the GFT basis of a DAG,  is also introduced using the eigenvector phases, as an alternative to the common total variation factor. The basic idea for the proposed concept was introduced in the special case of connected DAGs in \cite{telfor}.

We use the adjacency matrix as a suitable tool for the signal shift operator in directed graphs and apply it to construct graph Fourier transforms.   In classical (discrete-time domain) case, due to the distinctness of the eigenvalues of the adjacency matrix, a unique eigenvalue decomposition occurs, leading to the emergence of the discrete Fourier transform (on circular graphs).  From the linear algebra perspective, any graph Fourier transform provides a basis, functioning as harmonic vectors (with specified frequencies) and allowing the representation of a graph signal on a frequency-characterized basis. 
Several tools, employing different approaches, have been suggested to broaden the scope of spectral analysis of graph signals in both directed and undirected graphs.   
Here, we will mention two more related works in a similar line but with different approaches and final goals.  
Instead of commencing with the decomposition of the adjacency matrix as the graph descriptor, a novel method based on a specific form of the Laplacian for directed graphs is proposed in \cite{Sardellitti} to construct a graph Fourier basis that minimizes the total variation.
A similar approach to constructing a graph Fourier basis has been thoroughly investigated in \cite{Sevi}, where the eigenvectors of the random walk operator \cite{LS_Graph} are regarded as a non-orthogonal Fourier-type basis for representing graph signals. The frequency is interpreted by connecting the variation in these eigenvectors (obtained from their Dirichlet energy) to the real part of their associated eigenvalues.

Finally, we will emphasize that our final goal is to use the
adjacency matrix as a natural shift operator and to change
the graph in such a way that the GFT and spectral analysis
is possible, while the output of a graph system (graph filter)
on the altered graph is the same as if the graph were not
changed. This has not been considered or achieved in any
other paper.

The paper is organized as follows. In Section 2, the basic theory regarding graphs signals on a graph, systems on a graph, and the GFT is presented. Next, in Section 3 we present some fundamental properties of DAGs. Classical signal processing in a graph framework is presented in Section 4. In Section 5 we present graph zero-padding on a special case of connected DAG. General DAG zero-padding is presented in Section 6. Section 7 concludes the paper. The proposed approach is illustrated in numerical examples.

\section{Graph Signal Processing}

 Consider a graph $\mathcal{G}=(\mathcal{V},\mathcal{E})$ consisting of a finite set $\mathcal{V}$ of $N$ vertices, and set $\mathcal{E}\subset \mathcal{V} \times \mathcal{V}$ of edges, connecting the vertices, and reflecting their mutual relations. Each edge is represented by ordered pair of vertices $(m,n)\in \mathcal{V} \times \mathcal{V}$ indicating a starting vertex $m$ and an ending vertex $n$.
The graph can be modeled by the adjacency matrix $\mathbf{A}$. The elements $a_{mn}$ of this matrix have value $1$ if an edge between vertex $m$ and vertex $n$ exists, and value $0$ if such edge does not exist.

A graph signal is defined as an $N$-dimensional vector
\begin{equation}
 \mathbf{x}=
\begin{bmatrix}
x(1) & x(2) & \ldots & x(N)
\end{bmatrix}^T,
\end{equation}
where the signal value $x(n)$ is associated to the $n$-th vertex of the considered graph.
Graph shift operator is a matrix $\mathbf{T}$ that converts given graph signal $\mathbf{x}$ into its shifted version $\mathbf{y}$,
$$
\mathbf{y}=\mathbf{T}\mathbf{x}.
$$
The simplest and most commonly used shift operator on directed graphs is the adjacency matrix, i.e. $\mathbf{T}=\mathbf{A}$. That kind of shift is called a ``backward shift''.

A linear system on a graph is defined as a linear combination of the signal and its shifted versions. The output of a linear system on graph signal can be calculated as
\begin{equation}
 \mathbf{y}= h_0 \mathbf{x} + h_1 \mathbf{T} \mathbf{x} + h_2 \mathbf{T}^2 \mathbf{x} + \cdots + h_{S} \mathbf{T}^{S} \mathbf{x},
 \label{systemGA}
\end{equation}
where $S$ denotes the system order, and $h_s$, $s=0,1,\ldots,S$, are the system coefficients, corresponding to classical impulse response \cite{trends1,trends2,trends3}.

Note that every system on a given signal graph can be determined with $S<N$ coefficients, due to the fact that an arbitrary matrix power $\mathbf{T}^n$ can be expressed as a linear combination of matrix powers $\mathbf{T}^m$, $m=0,1,\ldots,N-1$.

The GFT is, for the case of directed graphs, defined with the assumption that the adjacency matrix is diagonalizable, meaning that there exists an invertible matrix $\mathbf{V}$ and a diagonal matrix $\mathbf{\Lambda}$ such that
$$
\mathbf{A}=\mathbf{V}\mathbf{\Lambda}\mathbf{V}^{-1}.
$$
Diagonal elements of matrix $\mathbf{\Lambda}$ are the eigenvalues, $\lambda_k$, $k=1,2,\dots,N$, of the adjacency matrix $\mathbf{A}$, whereas the columns of $\mathbf{V}$ are the eigenvectors, $\mathbf{v}_k$, $k = 1,2,\dots,N$.
The GFT of a signal, $x(n)$, on vertices $n=1,2,\dots,N$, is defined as
\begin{equation}
	X(k)=\sum_{n=1}^{N} x(n)u_k(n).
	\label{formulagft}
\end{equation}
Here, $u_k(n)$ represents the value of the $k$th column of matrix $\mathbf{V}^{-1}$, corresponding to the vertex $n$.
The matrix form of the GFT reads $
\mathbf{X}=\mathbf{V}^{-1} \mathbf{x}$.
The inverse graph Fourier transform (IGFT) is defined by 
\begin{gather}
	x(n)=\sum_{k=1}^{N} X(k)v_k(n),
	\label{formula_igft}
\end{gather}
or, in a matrix form,	$\mathbf{x}=\mathbf{V} \mathbf{X}$. Here, $v_k(n)$ is the value of the $k$th eigenvector (column of matrix $\mathbf{V}$), corresponding to the vertex $n$.

The system on a graph signal can be written using the GFTs of the input and output signal as \cite{trends2}
\begin{align}
	\mathbf{Y} & =  \big(h_0 \mathbf{I} + h_1 \boldsymbol{\Lambda}  + h_2 \boldsymbol{\Lambda}^2  + \cdots + h_{S} \boldsymbol{\Lambda}^{S}\Big) \mathbf{X},  \ \ \mathrm{ or  }\nonumber \\
	\mathbf{Y} & = \mathbf{H}(\boldsymbol{\Lambda}) \mathbf{X}, 
\end{align}
where $\mathbf{H}(\boldsymbol{\Lambda})$ is the system transfer function. 
\section{Directed Acyclic Graphs (DAG)}

\begin{figure}
	\centering
		
		\includegraphics[scale=0.75]{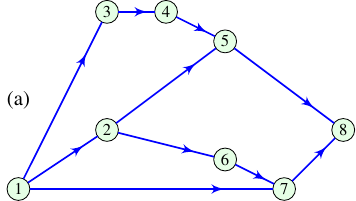} \hspace{5mm} 	\includegraphics[scale=0.75]{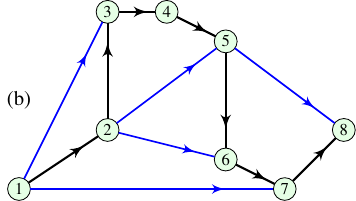}
	\caption{Examples of a directed acyclic graph (DAG): (a)  Disconnected DAG (for example, a path connecting vertices 2 and 3 does not exist), (b) Connected DAG (there exists a path between any pair of vertices).  A Hamiltonian path $1\to2\to3\to4\to5\to6\to 7\to 8$, which visits all vertices of the connected DAG exactly once, is denoted by black lines.}
	\label{DAG_E}
\end{figure}

In many cases of practical interest, especially for DAGs, the adjacency matrix is not diagonalizable and the Fourier analysis cannot be performed without graph modifications. Two examples  of a DAG with $N=7$ vertices are presented in Fig. \ref{DAG_E}. The graph at the top is not connected, and the graph at the bottom is connected.

Every DAG induces partial ordering on a vertex set. If vertex numbering follows this partial ordering, then the adjacency matrix will be an upper triangular matrix, with zeros on the main diagonal (since loops are not allowed in acyclic graphs).
 A connected DAG is a specific case of these graphs, where for each pair of vertices $(m,n)$ there exists a path from $m$ to $n$ or a path from $n$ to $m$.
For a connected DAG, the total ordering is achieved, as well as a unique numbering of graph vertices that corresponds to a Hamiltonian path  (a path that visits all vertices of a connected DAG exactly once), as shown in Fig. \ref{DAG_E}(b).   The existence of a Hamiltonian path in a DAG enforces a specific topology on the edge directions. All edges align with the Hamiltonian path direction, leading to the uniqueness of the Hamiltonian path.  The uniqueness of the Hamiltonian path implies that two connected DAGs with different adjacency matrices (in upper triangular form) are non-isomorphic. This further indicates that the total number of connected DAGs with $N$ vertices  is 
\begin{equation}
\label{eq:ndag}
N_{\mathrm{ConDAG}}=2^{\frac{N(N-3)}{2}+1}.
\end{equation}
We can derive this formula starting from the adjacency matrix of connected DAG, with nodes numbered according to their position in the Hamiltonian path, which have ones at the super-diagonal. Above the super-diagonal, we have a total of ${N(N-3)}/{2}+1$ elements that could be either 1 or 0, so the total number of connected DAGs directly follows.

\noindent\textbf{Remark:} Adjacency matrix of a DAG is nilpotent with nilpotency index smaller than $N$.
 This means that there exists a nilpotency index $m$, such that $\mathbf{A}^{m}=\mathbf{0}$ and $\mathbf{A}^{m-1}\ne\mathbf{0}$.
This further implies that all eigenvalues of a DAG adjacency matrix are equal to 0.

If the adjacency matrix, $\mathbf{A} $, is used as a shift operator on DAGs, then any system on a DAG can be written in the following form
\begin{equation}
 \mathbf{y}= h_0 \mathbf{x} + h_1 \mathbf{A} \mathbf{x} + h_2 \mathbf{A}^2 \mathbf{x} + \cdots + h_{S} \mathbf{A}^{S} \mathbf{x},
\end{equation}
where $S+1$ is the nilpotency index of $\mathbf{A}$.

\section{Classical Zero-Padding within the Graph Framework}

\begin{figure}
\centering
		\includegraphics[scale=0.75]{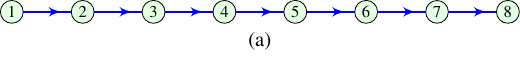}\\ \vspace{5mm}
		\includegraphics[scale=0.75]{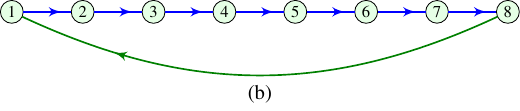}\\  \vspace{5mm}
		\includegraphics[scale=0.75]{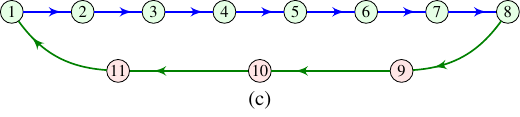}
		
	\caption{(a) The domain of classical discrete-time signal of length $N=8$. (b) The domain of classical signal commonly used in DFT calculations with $N=8$ (for example, as in \cite{seifert2021digraph}). (c) The domain of the zero-padded classical signal required for DFT-based calculations of the output of an FIR system order $M\leq 3$, with $N=8$.}
	\label{DAG_123}
\end{figure}

If we have a classical signal in the discrete-time domain, $x(n)$, of length $N$, then its domain can be presented in the form of a directed path graph with $N$ vertices, as shown in Fig. \ref{DAG_123}(a). 

 \noindent \textbf{Remark:} It is well known that for the calculation of the discrete Fourier transform (DFT) of this signal we must assume signal periodicity. In graph terminology, it means that we have to make the graph domain of the signal in a circular manner. The minimum number of samples for the DFT calculation of the assumed signal is $N$, meaning that we inherently add one edge connecting the last and the first vertex, see Fig. \ref{DAG_123}(b). 
 
 This necessary and obvious domain modification can be interpreted and justified within the graph terminology as well. The adjacency matrix, $\mathbf{A}$, of the path graph from Fig. \ref{DAG_123}(a) is a super-diagonal matrix
 \begin{equation*}
 	\mathbf{A}=\begin{bmatrix}
 		0 & 1 & 0 & 0 & 0 & 0 & 0 & 0\\
 		0 & 0 & 1 & 0 & 0 & 0 & 0 & 0\\
 		0 & 0 & 0 & 1 & 0 & 0 & 0 & 0\\
 		0 & 0 & 0 & 0 & 1 & 0 & 0 & 0\\
 		0 & 0 & 0 & 0 & 0 & 1 & 0 & 0\\
 		0 & 0 & 0 & 0 & 0 & 0 & 1 & 0\\
 		0 & 0 & 0 & 0 & 0 & 0 & 0 & 1\\
 		0 & 0 & 0 & 0 & 0 & 0 & 0 & 0\\
 	\end{bmatrix},
 \end{equation*}
 with all eigenvalues following from $\mathrm{det}\big(\mathbf{A}-\lambda \mathbf{I}\big)=0$, being $$\lambda_k=0, \ \text{ for } \ k=1,2,\dots,N.$$ 
 This adjacency matrix $\mathbf{A}$ is not diagonalizable and the Fourier analysis on this graph (acting as the signal domain) is not possible). Note that, if we assume an undirected graph as the signal domain, then the Fourier analysis would be possible. Such an assumption, however, completely changes the physics of the problem, and is, therefore, not appropriate in this context. 
 
  \noindent \textbf{Diagonalization by making the domain circular.}  If we make the domain circular, by adding one edge from the last (sink) vertex to the first (source) vertex, we get the adjacency matrix 
 \begin{equation}
 	\mathbf{A}_C=\begin{bmatrix}
 		0 & 1 & 0 & 0 & 0 & 0 & 0 & 0\\
 		0 & 0 & 1 & 0 & 0 & 0 & 0 & 0\\
 		0 & 0 & 0 & 1 & 0 & 0 & 0 & 0\\
 		0 & 0 & 0 & 0 & 1 & 0 & 0 & 0\\
 		0 & 0 & 0 & 0 & 0 & 1 & 0 & 0\\
 		0 & 0 & 0 & 0 & 0 & 0 & 1 & 0\\
 		0 & 0 & 0 & 0 & 0 & 0 & 0 & 1\\
 		1 & 0 & 0 & 0 & 0 & 0 & 0 & 0\\
 	\end{bmatrix}. \label{AcMat}
 \end{equation}
 This matrix is obviously diagonizable. Eigenvalues of matrix $\mathbf{A}_C$,  of size $N\times N$, are obtained as the solutions of
 $$ \lambda^N=1.$$
 They are all distinct and equal to
 $$ \lambda_k=e^{j\frac{2\pi}{N}(k-1)}, \quad k=1,2,\ldots,N,$$
 where $k$ is the frequency index. 
 
 \noindent \textbf{Frequency interpretation.} For $1\le k \le \frac{N}{2}$ we talk about ``positive'' frequencies, while indices $\frac{N}{2} <  k \le N $ are associated to  ``negative'' frequencies. Therefore, the graph spectral analysis on a circular graph coincides with classical DFT analysis. 
 Note that the discrete frequency $\omega_k=\frac{2\pi}{N}(k-1)$ can be obtained as an angle of the corresponding eigenvalue $\lambda_k$, that is, $$\omega_k=\arg \{\lambda_k\}.$$  
Note that, in classical DFT, the frequency indicates the speed of change in the eigenvectors. Another way to measure the changes in the eigenvectors (with GFT basis)  is by using the total variation \cite{trends2,ChenS}, i.e.
 $$
 E_{TV}(k)=\Big|1-\frac{\lambda_k}{\max_m | \lambda_m | }\Big|.
 $$
This parameter is used in graph signal processing (especially for undirected graphs), to sort the eigenvectors from low to high frequency, but it cannot distinguish ``positive'' and ``negative'' frequencies. 
  
\begin{figure}
\centering
		\includegraphics[scale=0.75]{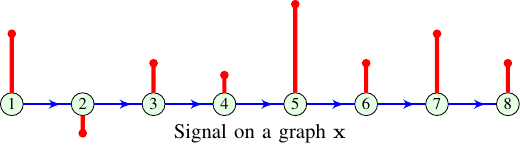}\\ \vspace{2mm}
		\includegraphics[scale=0.75]{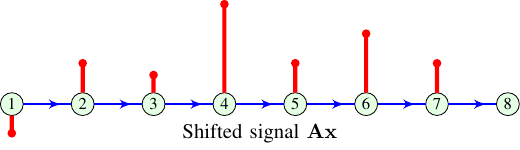}\\  \vspace{2mm}
		\includegraphics[scale=0.75]{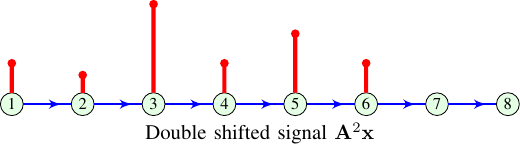}\\ \vspace{2mm}
		\includegraphics[scale=0.75]{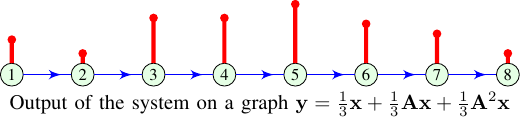}
	\caption{Signal $x(n)$  of length $N=8$ on a graph as its domain (first row), along with shifted versions of the same signal (second and third rows). The last sub-figure represents the output of a system on a graph of order $S=2$ with system coefficients equal to $1/3$.}
	\label{DAG_x}
\end{figure}

\noindent \textbf{Signal processing and zero-padding.} Assume now that we want to process the considered classical signal $x(n)$ by a general FIR system whose impulse response $h_n$ is of  length $(S+1)$, that is, by a system of order $S$. This classical system is defined within the graph framework by (\ref{systemGA}). Its discrete-time domain realization can be written as
 $$y(n)=h_0x(n)+h_1x(n-1)+\dots+h_{S}x(n-S).$$ 
 If we want to use the DFT to calculate the output (which is almost a standard routine in DSP processors), then we must zero-pad the original signal by at least $M=S$ zeros before the DFT is applied (as illustrated in Fig. \ref{DAG_123}(c)) , \cite[p. 591]{DSPbook1}, \cite{DSPbook}. After zero-padding, the DFT will produce the output that corresponds to the original non-periodic output signal that is shown in Fig. \ref{DAG_x}. 
 
\begin{figure}
\centering
{\footnotesize{(a)}}
		\includegraphics[scale=0.75]{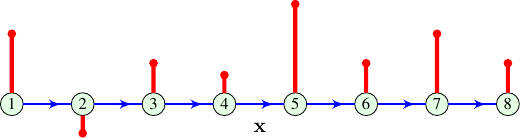}\\
{\footnotesize{(b)}} \vspace{2mm}
		\includegraphics[scale=0.75]{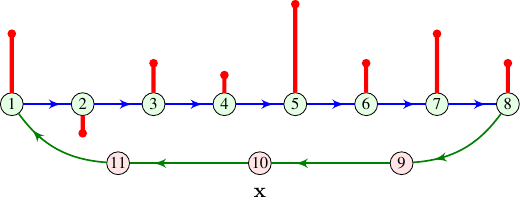}\\
	{\footnotesize{(c)}}	
	 \vspace{2mm}
		\includegraphics[scale=0.75]{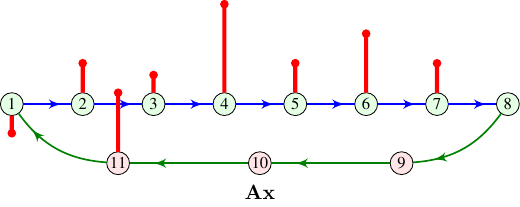}\\
		{\footnotesize{(d)}}
			 \vspace{2mm}
		\includegraphics[scale=0.75]{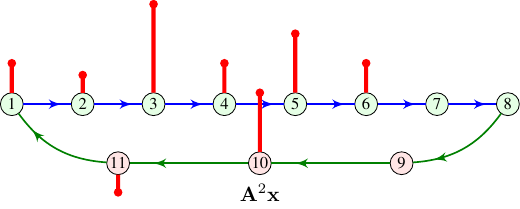}\\
		{\footnotesize{(e)}} \vspace{2mm}
		\includegraphics[scale=0.75]{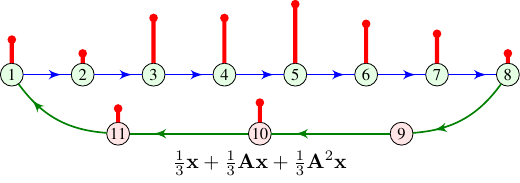}	
	\caption{(a) Signal $x(n)$  of length $N=8$ on a graph as its domain. (b) Signal on a graph as it must be used in the DFT output calculation on a FIR system whose order is $S \le 3$. (c) Signal $x(n-1)$ shifted on a graph from (b). (d) Signal $x(n-2)$ on a graph from (b). (e) Output signal as a circular (DFT) convolution $\mathbf{y}=\mathbf{x}*[\frac{1}{3},\frac{1}{3},\frac{1}{3}]$ equal (within the basic period) to the aperiodic convolution of the same signal and system. }
	\label{DAG_456}
\end{figure}

 Within the graph terminology, the zero-padding of the original signal (before we make it periodic) means that we must add at least $M$ vertices in the backward path from the sink to the source vertex, as illustrated in Fig. \ref{DAG_456}. The adjacency matrix of the zero-padded graph is of the same form as in (\ref{AcMat}), but of order $N+M$. It is diagonalizable with all distinct  eigenvalues 
 $$\lambda_k=e^{j\frac{2\pi}{N+M}(k-1)}, \quad k=1,2,\ldots,N+M,$$
 and corresponds to the DFT analysis of a signal with $N+M$ samples, after zero-padding.

\vspace{-1mm}

\section{Connected DAG Zero-Padding}
Connected DAG has one source vertex and one sink vertex. Moreover, there exists a path from sink to source passing through every vertex in a DAG (Hamiltonian path).

Based on the discussion in the previous section, we can conclude that a possible solution for a non-diagonalizable adjacency matrix is to add an edge from the sink to the source vertex in order to obtain a new graph with a diagonalizable adjacency matrix. If we want to keep the same output of a graph filter (of an order less than or equal to $M$) on the original and a modified DAG, then instead of adding an edge we should add a path of length $M$ connecting the sink and the source in the original connected DAG.  

With the described zero-padding, the shifts produced by $\mathbf{A}^{m} \mathbf{x}$, $m=0,1,2,\dots,M$  on the original graph, will correspond to the shifts on the zero-padded graph, while the shifted signal values, will be trapped in the added path from the sink to the source that is long enough so they do not influence the signal at the vertices that belong to original graph, during the system processing time. 

The adjacency matrix of the zero-padded graph can be presented in a block form as
\begin{equation}
	\mathbf{A}_{ZP}=
	\begin{bmatrix}
		\mathbf{A} & \mathbf{C} \\
		\mathbf{D} & \mathbf{J}
	\end{bmatrix},
\end{equation}
where $\mathbf{A}$ is $N\times N$ adjacency matrix of the connected DAG, $\mathbf{J}$ is a matrix of size $M\times M$, with ones at super-diagonal and zeros elsewhere. Matrix $\mathbf{C}$ is of size $N\times M$, with all zeros except the lower left corner, where it has value 1. This block represents the connection of the DAG sink to the first vertex of the added path graph. Matrix $\mathbf{D}$ is of size $M\times N$ with all zeros except the lower left corner where it has value 1. This block represents the edge from the last vertex of the added path graph to the source of the DAG.

\noindent \textbf{Determinant in the zero-padded graph:} It is easy to check that determinant of $\mathbf{A}_{ZP}$ is non-zero. To be more specific, its values are either 1 or $-1$.

\begin{figure}
\centering
	\includegraphics[scale=0.75]{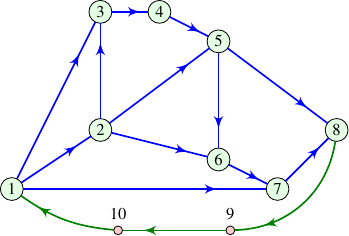}
	\caption{An illustration of a zero-padded DAG. The DAG shown in Fig. \ref{DAG_E} (b) is zero-padded with additional vertices (marked in red) and edges (highlighted in green), necessary for the GFT-based analysis.}
	\label{DAG_FZP}
\end{figure}

To illustrate the proposed concept, consider a connected DAG with $N=8$ vertices (Fig. \ref{DAG_E}(b)), with the adjacency matrix   
\begin{equation*}
	\mathbf{A}= \begin{bmatrix} 
		0 & 1 & 1 & 0 & 0 & 0 & 1 & 0\\
		0 & 0 & 1 & 0 & 1 & 1 & 0 & 0\\
		0 & 0 & 0 & 1 & 0 & 0 & 0 & 0\\
		0 & 0 & 0 & 0 & 1 & 0 & 0 & 0\\
		0 & 0 & 0 & 0 & 0 & 1 & 0 & 1\\
		0 & 0 & 0 & 0 & 0 & 0 & 1 & 0\\
		0 & 0 & 0 & 0 & 0 & 0 & 0 & 1\\
		0 & 0 & 0 & 0 & 0 & 0 & 0 & 0
	\end{bmatrix}.
\end{equation*}
For zero-padding with $M=2$ additional vertices, the matrix $\mathbf{A}_{ZP}$ has the following form
\begin{equation*}
	\mathbf{A}_{ZP}=\left[\begin{array}{cccccccc|cc}
	0 & 1 & 1 & 0 & 0 & 0 & 1 & 0 & 0 & 0\\
	0 & 0 & 1 & 0 & 1 & 1 & 0 & 0 & 0 & 0\\
	0 & 0 & 0 & 1 & 0 & 0 & 0 & 0 & 0 & 0 \\
	0 & 0 & 0 & 0 & 1 & 0 & 0 & 0 & 0 & 0\\
	0 & 0 & 0 & 0 & 0 & 1 & 0 & 1 & 0 & 0\\
	0 & 0 & 0 & 0 & 0 & 0 & 1 & 0 & 0 & 0\\
	0 & 0 & 0 & 0 & 0 & 0 & 0 & 1 & 0 & 0\\
	0 & 0 & 0 & 0 & 0 & 0 & 0 & 0 & 1 & 0 
	\\ \hline
	0 & 0 & 0 & 0 & 0 & 0 & 0 & 0 & 0 & 1\\
	1 & 0 & 0 & 0 & 0 & 0 & 0 & 0 & 0 & 0 
	\end{array}\right],
\end{equation*}
and we obtain a zero-padded graph presented in Fig.  \ref{DAG_FZP}.
This adjacency matrix is now diagonizable and a system on this DAG can be analyzed and implemented in the spectral domain. 

\subsection{Numerical Examples}

\begin{figure}[ptb]
	\centering
	\includegraphics[width=12cm]{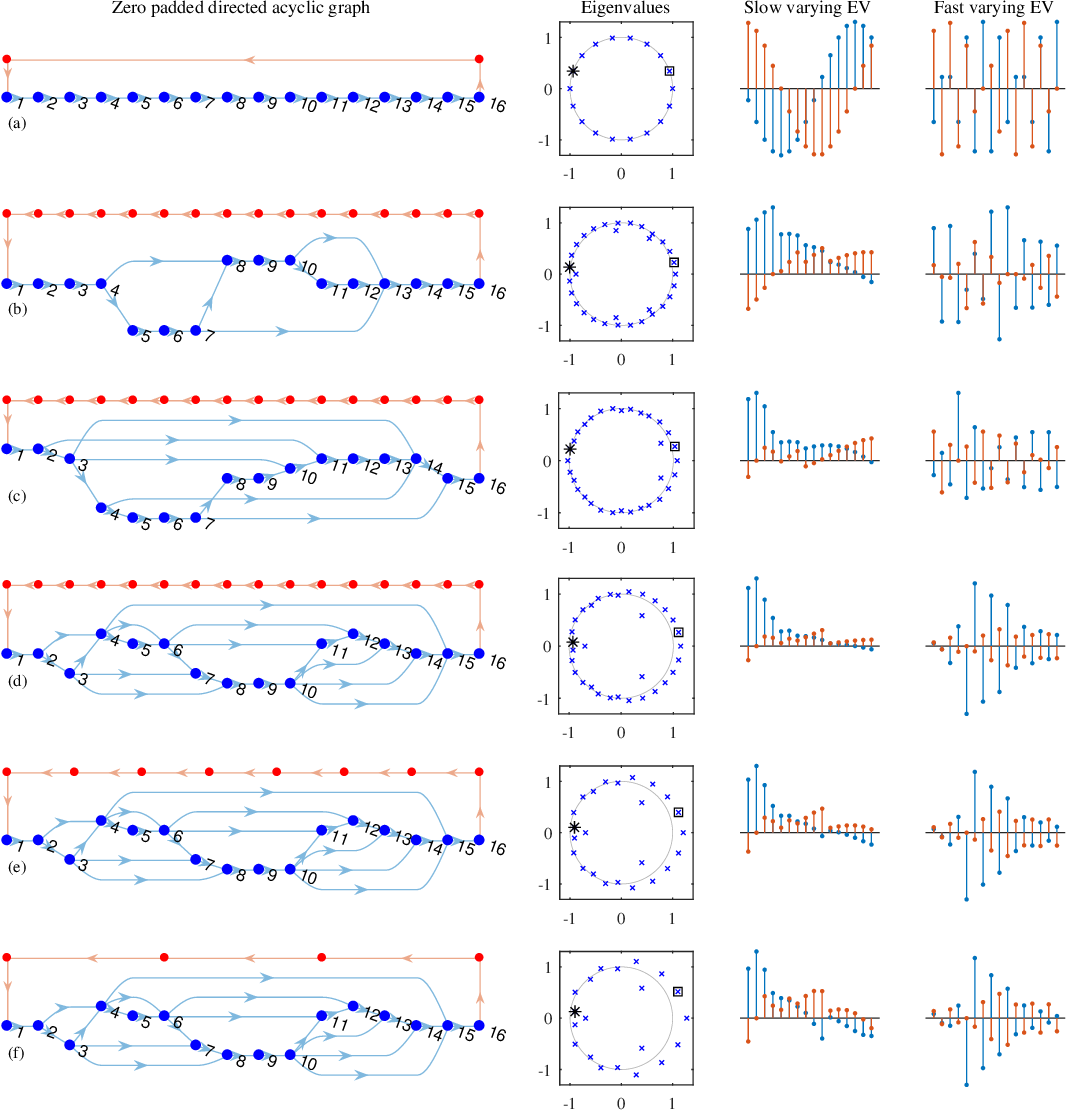}
	\caption{Examples of zero-padded connected DAGs. Simple path graph (a) zero-padded with $M=2$ vertices; connected DAGs with 3, 5, and 10 additional edges compared to the path graph from (a), and zero-padded with $M=16$ vertices (b-d); the same DAG as in (d) zero-padded with $M=8$ and $M=4$ (e,f).
		Original DAGs are presented with blue vertices and edges. Nodes and edges added by zero-padding are colored in red. 
		For each zero-padded graph eigenvalues are presented. Two eigenvectors that correspond to a low and to a high frequency eigenvalue are also presented (real parts in blue and imaginary parts in red). The corresponding eigenvalues are marked by a square (low-frequency) and a star (high-frequency) on the eigenvalues plot. }
	\label{gzp_ex4_var}
\end{figure}

Five examples of zero-padded connected DAGs are presented in Fig. \ref{gzp_ex4_var}. The first case (a path graph) is the simplest. This graph transforms into a cycle graph  after  zero-padding (here we use zero-padding with $M=2$), shown in Fig. \ref{gzp_ex4_var}(a). The classical Fourier analysis follows straightforwardly from  the eigendecomposition of the adjacency matrix. Eigenvalues are equally spaced on the unit circle, as shown in Fig. \ref{gzp_ex4_var} (second column).

In more complex DAG examples, as in Fig. \ref{gzp_ex4_var}(b,c,d),  more edges are added to the path graph in order to keep the connectivity and avoid cycles. Here, we present graphs with 3, 5, and 10 additional edges. In all of these cases, we can see from Fig. \ref{gzp_ex4_var} that the eigenvalues lie close to the unit circle, with nonuniform, but almost equal spacing. Here, graph zero-padding with $M=N=16$ is used.
Finally in Fig. \ref{gzp_ex4_var}(e,f), the same DAG as in  Fig. \ref{gzp_ex4_var}(d) is zero-padded with $M=8$ and $M=4$ vertices, respectively. Using zero-padding with $M<N$ guarantees equality of the output signal calculated in the vertex domain and output calculated using the spectral domain for systems of order up to $M$.

For each considered graph in Fig. \ref{gzp_ex4_var}, two eigenvectors are presented on the right. Since eigenvectors are complex, we present their real part in blue and the imaginary part in red. We choose one slow-varying eigenvector that corresponds to a low frequency  (marked by a square in the eigenvalue plot in Fig. \ref{gzp_ex4_var}) and one fast-varying eigenvector with the corresponding eigenvalue marked by a star. From these plots, we observe that the concept of slow-varying and fast-varying Fourier basis functions (eigenvectors), obvious in the classical Fourier domain (circular graph), 
remains unchanged in the considered case of GFT defined on a zero-padded DAG.

\section{General DAG zero-padding}

For a general DAG, we should consider adding more than one edge (or path, if zero-padding is applied) to the original DAG. This step ensures that the adjacency matrix of the new graph becomes diagonalizable, with all non-zero eigenvalues, making the GFT analysis possible.
Determining how to add the minimal number of edges to make the adjacency matrix diagonalizable is not an easy task. It can be solved using a combinatorial approach (exhaustive search), but only for a small graph size $N$. Another recently proposed approach uses matrix perturbation theory \cite{seifert2021digraph}. Here, we will propose a simple and fast, albeit  sub-optimal strategy. It can be presented in two steps:
\begin{itemize}
\item First, we add a sufficient number of edges in order to make the DAG connected, while preserving the acyclic nature of the graph. We can use the procedure described in Algorithm \ref{alg-dag-connecting}. 
 Note that if we intend to implement zero-padding, then each edge added in Algorithm \ref{alg-dag-connecting} should be replaced by a path of length $M+1$, incorporating $M$ additional vertices along it.
\item Secondly, we add an edge connecting sink and source of the connected DAG, obtained in the previous step.
\end{itemize}

Within Algorithm \ref{alg-dag-connecting}, we aim to make the DAG connected by establishing connections between existing vertices. 

This procedure is performed directly, meaning that we search for all sources in a DAG, then proceed with determining the edges that will link these sources, subsequently removing these sources from the DAG. By repeating this process, we ensure that a sufficient number of edges is added to achieve connectivity. At the end, we find a Hamiltonian path in the new (connected) DAG and identify which edges from the previous procedure belong to the Hamiltonian path. These edges are essential for achieving the connectivity of the DAG. Any added edges outside the Hamiltonian path can safely be removed.

The while loop in Algorithm \ref{alg-dag-connecting} will be executed at most $N$ times, since at least one vertex is removed from $\mathcal{G}_1$ in each iteration. Within the while loop, identifying all sources in $\mathcal{G}_1$ requires $O(N^2)$ operations. The total computational complexity of Algorithm \ref{alg-dag-connecting} is $O(N^3)$.

If we renumber vertices in the connected DAG according to their position in the Hamiltonian path, the  adjacency matrix becomes upper triangular, with ones on the super-diagonal. Adding vertices in the return path with an appropriate numbering ($\text{sink}+1,\text{sink}+2,\ldots$) keeps ones at the adjacency matrix super-diagonal and adds 1 into its lower left corner.

Denote by $K$ the number of edges needed to make the DAG connected.
If we zero-pad each added edge, replacing it with a path consisting of $M$ vertices, the size of the obtained adjacency matrix of the connected zero-padded DAG increases to $N+KM$. Adding a zero-padded return path additionally increases the adjacency matrix size to $N+(K+1)M$.

\begin{algorithm}[tb]
	\caption{Establishing Connectivity in a DAG}
	\label{alg-dag-connecting}
	\begin{algorithmic}
		\Require Directed acyclic graph $\mathcal{G}$
		\Ensure Connected directed acyclic graph $\mathcal{G}_2$
		\State $\mathcal{G}_1 \gets  \mathcal{G}$
		\State $\mathcal{G}_2 \gets \mathcal{G}$
		\While{$\mathcal{G}_1$ is not empty}
		\State $s \gets $ list of all sources in $\mathcal{G}_1$
		\If{there is more than one source in $s$}
		\State connect sources in  $s$ by a path in $\mathcal{G}_2$
		\Else
		\State remove source $s$ from $\mathcal{G}_1$
		\EndIf
		\EndWhile
		\State $h \gets$ Hamiltonian path in $\mathcal{G}_2$
		\State remove all edges from $\mathcal{G}_2$ that are not in $\mathcal{G}$ and do not belong to $h$
		\State \Return  $\mathcal{G}_2$
	\end{algorithmic}
\end{algorithm}

	\begin{table}[tbp]
\centering
		\caption{Total number of connected DAGs ,  number of DAGs, among them, whose eigenvalues of the adjacency matrix are all distinct  after adding sink-to-source edge, and the number of such DAGs with an algebraic multiplicity of the adjacency matrix eigenvalue greater than one.}
		\label{Tab1}

			\begin{tabular}{lrrrr}
				\toprule
				& Total DAGs   & Diagonal. & \multicolumn{2}{c|}{Algeb. Mult.$\mathrm{}>1$}  \\ \midrule
				$N=7$ & $32,768$  & $32,250$  &  $518 $ & $1.58\%$ \\
				$N=7$ (1-ZP) & $32,768$ & $32,758$ &  $ 10 $ & $0.03\%$   \\
				$N=8$  &  $2,097,152$ & $2,075,682$ &  $21,470$ & $1.02\%$   \\
				$N=8$ (1-ZP) & $2,097,152$ &  $2,088,106$ &  $9,046$  & $0.43\%$ \\
				$N=8$ (2-ZP) & $2,097,152$ &  $2,095,224$ &  $1,928$ &  $0.09\%$ \\
				\bottomrule
			\end{tabular}  
	\end{table}

\noindent\textbf{Invertibility, Diagonalizability, and  Algorithm's Efficacy.}  {After Algorithm 1 is executed, the adjacency matrix attains full rank, indicating that it becomes invertible. This property implies that all eigenvalues of the considered matrix are non-zero.  However, the invertibility does not guarantee diagonalizability.}
{ The nondiagonalizability of the adjacency matrix of a graph obtained  after Algorithm 1 is a very rare event. Even in these rare cases, after the added edges are zero-padded, it often happens that the resulting adjacency matrix becomes  diagonalizable.

For relatively small graphs, with, for example, $N=7$ and $N=8$, it is feasible to construct all possible connected DAGs, with added sink-to-source connection (edge or path), and to examine the algebraic multiplicity of the eigenvalues. Having all distinct eigenvalues is sufficient for the diagonalizability. By adding only sink-to-source edge to all considered DAGs, the diagonalizability is achieved in 98,42\% and 99\% for $N=7$ and $N=8$, respectively.
These results are presented in Table \ref{Tab1}, rows one and three. 

After adding one or two vertices to the sink-to-source path, we observe that the adjacency matrix's diagonalizability increases to above 99.6\% of cases (rows two, four, and five in Table \ref{Tab1}).

It is very interesting that all 518  cases (out of 32768) that were not diagonalizable with  $N=7$, become diagonalizable after adding one vertex in the sink-to-source path. However, 10 graphs that were diagonalizable after the sink-to-source edge is added, become nondiagonalizable after this vertex is added. We checked that in all 32768 possible connected DAGs with $N=7$, and in all 2097152 possible connected DAGs for $N=8$, the diagonalizability is achieved either with a single sink-to-source edge or with one added vertex in the sink-to-source path. Since the number of added vertices, could be changed, as far as their number is greater or equal to the order of a system on the graph, it means that \textit{in this way the diagonalizability can be achieved, in  practically  (almost) all cases.} }

\textbf{Weighted DAG case.} The proposed approach can be directly extended to weighted DAGs (including the case of a random walk  matrix used as shift in the system). The only modification required is to assign weights to the added edges.

In this manner, addressing the diagonalizability issue becomes even simpler, as we have flexibility in choosing the weights for the added edges. For instance, in connected DAGs of size $N=7$ and $N=8$, there are cases where adding a single unweighted edge is insufficient for achieving diagonalizability (refer to Table \ref{Tab1}). However, by introducing a weighted edge with a weight of 0.5, all instances become diagonalizable.   

\subsection{Graph Zero-Padding from a Linear Algebraic Perspective}
In the initial step, it becomes clear that the nilpotency of the adjacency matrix, with nilpotency index $P$,  indicates that the geometrical multiplicity of the unique eigenvalue $0$ is exactly $P+1$. This, in turn, implies that in the Jordan normal form, there are precisely $P+1$ nontrivial Jordan blocks associated with the eigenvalue $0$. To achieve diagonalizability, it is imperative to eliminate all of these blocks. Algorithm \ref{alg-dag-connecting} addresses the task of consolidating all these non-trivial Jordan blocks into a single block while maintaining the acyclic nature of the graph.

After successfully constructing the Hamiltonian path, an extra edge is added to connect the sink to  the source. This requires the application of a new perturbation, which is in the form of a rank-one elementary matrix. 
Empirical evidence confirms that, in almost all scenarios, diagonalizability is successfully achieved in this phase, especially after zero-padding and adding additional vertices. 

In exceptional cases, the second perturbation may introduce a few nontrivial Jordan blocks, which can also be resolved by the inclusion of a small number of randomly positioned edges. Through these insightful observations and a careful examination of Algorithm \ref{alg-dag-connecting} on graphs, significant progress can be achieved by addressing the challenges in the way discussed in the recent related work \cite{seifert2021digraph}.

 The non-singularity of the GFT leads us to derive a boundary condition \cite{seifert2021digraph} when signals are presented in the polynomial ($z$-transform) representation. Suppose that the GFT is of size $N$. It can be considered as a polynomial transform that maps the $N$-dimensional space of polynomials, $\mathbb{C}_{N-1}[z]$, to $\mathbb{C}^N$. The Chinese Remainder Theorem implies the existence of a polynomial $p(z)$ of a degree $N$, such that it establishes an isomorphism between the quotient ring $\mathbb{C}[z]/p(z)$ and $\mathbb{C}^N$. The polynomial $p(z)$ describes the boundary condition for the considered case.

\subsection{Examples}
\begin{figure}[tbp]
\centering
	\includegraphics[scale=0.75]{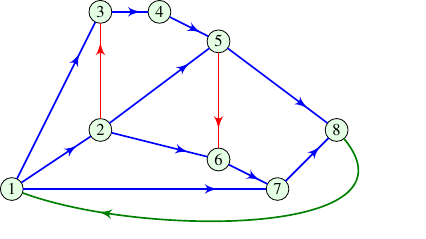}
	\caption{An example of a DAG with additional edges needed for the GFT analysis. Red edges are used to make the DAG connected with a Hamiltonian path 1-2-3-4-5-6-7-8, while the green edge is used for a sink-to-source edge, used to convert the Hamiltonian path to a Hamiltonian cycle.}
	\label{DAG_G}
\end{figure}

For the DAG presented in Fig \ref{DAG_E}(a), we should add two edges in order to produce a connected DAG (for example, edges $(2,3)$ and $(5,6)$). These edges are shown in red color in Fig. \ref{DAG_G}. By adding the return edge $(8,1)$, indicated by the green color in Fig. \ref{DAG_G}, we obtain the graph where we can define the GFT.

\begin{figure}[tbp]
\centering
	\includegraphics[scale=0.75]{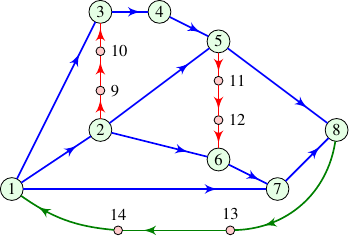} \\ \vspace{3mm}
	
		\includegraphics[scale=0.75]{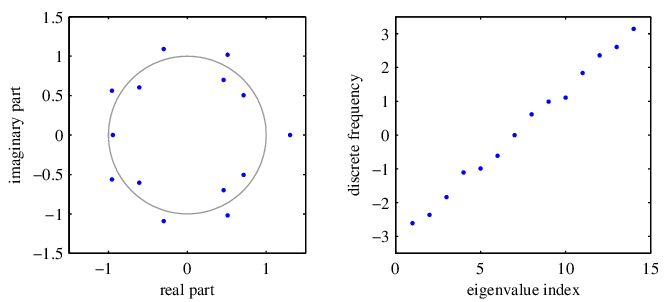}
		
	\caption{(top) An example of a zero-padded graph with additional vertices needed for achieving the same behavior of systems of order less than or equal to $S=2$ on the original DAG and the obtained zero-padded graph. (bottom) Eigenvalues of the zero-padded graph, with a relationship between the eigenvalue index and the corresponding discrete frequency.}
	\label{DAG_F}
\end{figure}

If we want to keep the same behavior of the original DAG and the modified DAG for signal processing and systems on graphs of maximal order $M$, we should insert new paths in the graph with additional $M$ vertices, instead of each edge which is added in the previous procedure. For the DAG presented in Fig. \ref{DAG_E}(a), by  using $M=2$ and introducing new vertices, the obtained zero-padded graph is presented in Fig. \ref{DAG_F} (top).

\begin{figure}[tbp]
	\centering
	\includegraphics[scale=0.75]{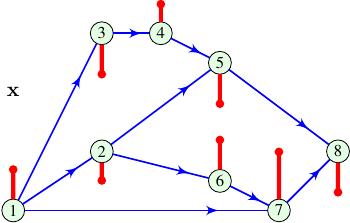}
	\includegraphics[scale=0.75]{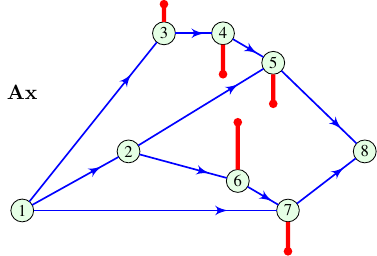}
	\includegraphics[scale=0.75]{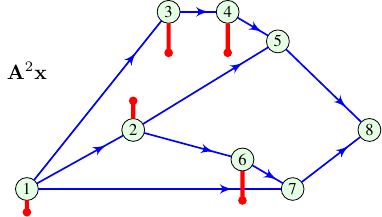}
	\includegraphics[scale=0.75]{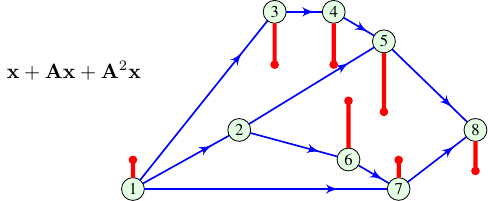}
	\caption{An example of signal processing using the second order system on a DAG.}
	\label{DAG_ESig}
\end{figure}

\begin{figure}[tbp]
	\centering
	\includegraphics[scale=0.75]{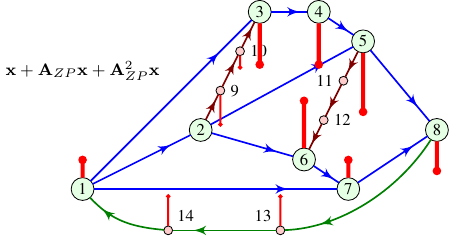}
	\caption{ Signal processing on the DAG from Fig. \ref{DAG_ESig}, after zero-padding is done, so that the GFT can be used.}
	\label{DAG_FSist}
\end{figure}

In the considered case, we can add more than $M$ zero-padding vertices in the added paths (sometimes $M$ is not known in advance). Having in mind that the adjacency matrix of the  original DAG, shown in Fig. \ref{DAG_E}(a), has a nilpotency index of 5, meaning that any system on that graph is of order smaller than 5. By inserting $M=4$ vertices in each path, an arbitrary system defined on the original DAG and applied to the zero-padded DAG will produce the same output (when we neglect signal values at the vertices added through the zero-padding procedure).

As an example, consider a signal on the DAG presented in Fig. \ref{DAG_ESig} (top). Shifted signals are presented in the same Figure. When we apply a system of the second order on this graph, defined by
$$
\mathbf{y}=\mathbf{x}+ \mathbf{A}\mathbf{x}+\mathbf{A}^2\mathbf{x},
$$
the output signal is given in Fig. \ref{DAG_ESig} (bottom).

The same signal and system are then used on a zero-padded DAG shown in Fig. \ref{DAG_F} (top), assuming zero signal values at the added vertices ($9,10,\ldots,14$). The system output is presented in Fig. \ref{DAG_FSist}. The output signal at vertices ($1,2,\ldots,8$) is the same as in the case of the original DAG (see Fig. \ref{DAG_ESig} (bottom)). Signal values at the vertices added to the original DAG by zero-padding procedure could be discarded.

\begin{figure}[tbp]
\centering
\includegraphics[width=\linewidth]{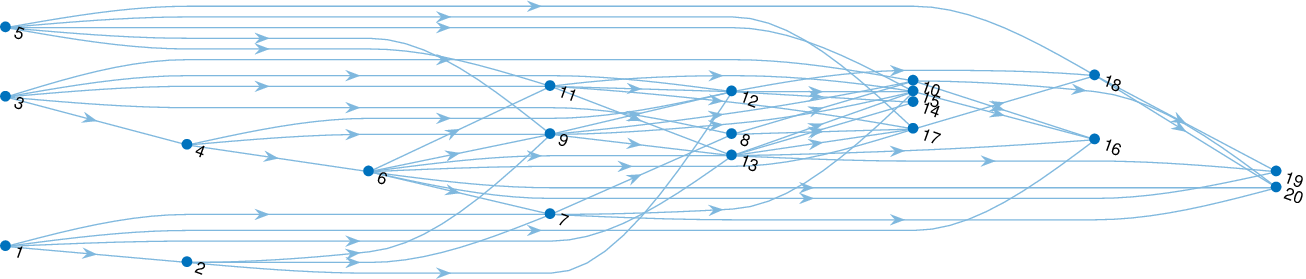}\\(a)\\[2ex]

\includegraphics[width=\linewidth]{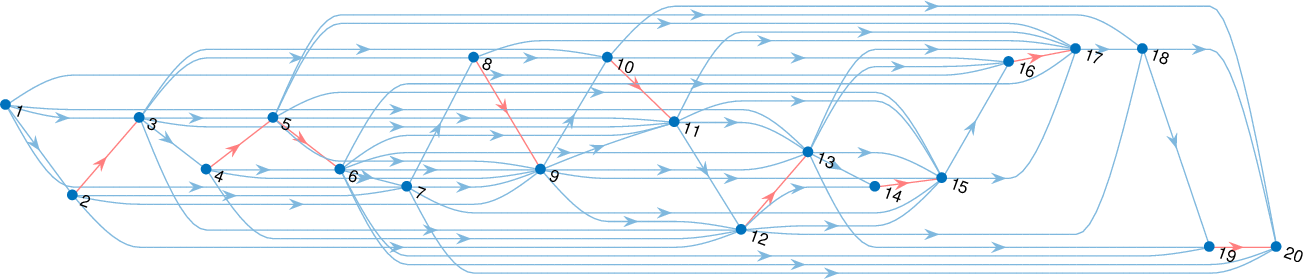}\\(b)\\[2ex]

\includegraphics[width=\linewidth]{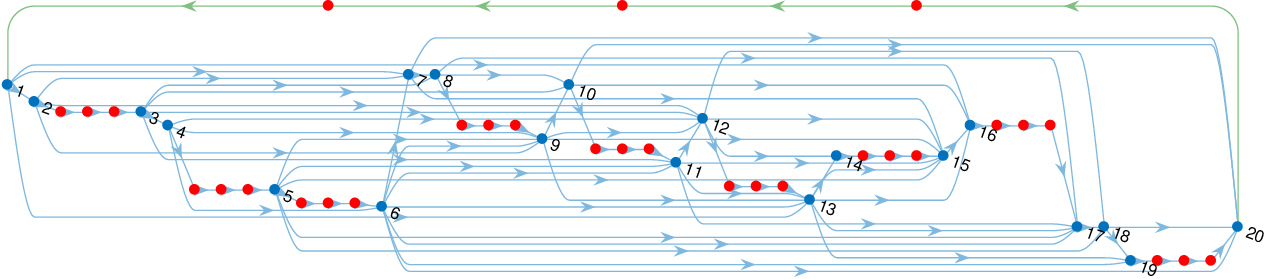}\\(c)\\[2ex]

\caption{(a) Original DAG, (b) connected DAG with added edges marked by red color, (c) zero-padded DAG with added vertices ($M=3$) marked by red color and return path marked by green color.}
\label{dag_ex2}
\end{figure}

\begin{figure}[tbp]
\centering
\includegraphics[scale=0.75]{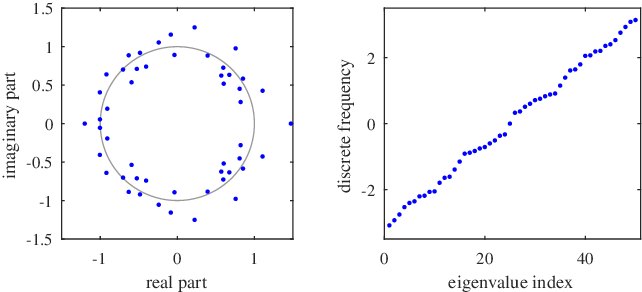}
\caption{Eigenvalues of the zero-padded connected DAG shown in Fig. \ref{dag_ex2}(c) (left); the relation between the eigenvalue index and the discrete frequency value (right).}
\label{dag_ex2_ev}
\end{figure}

A more complex example is shown in Fig. \ref{dag_ex2}(a). Here, we consider a DAG with $N=20$ vertices and 54 edges. There are three sources ($1,3$ and $5$) and two sinks ($19$ and $20$). This DAG is obviously not connected. Next, we add edges, according to Algorithm \ref{alg-dag-connecting}, in order to make the DAG connected. Added edges are shown in red color in Fig. \ref{dag_ex2}(b). Then, we perform zero-padding by replacing each added edge by a path graph with $M=3$ additional vertices, for each path, and adding a return path from the sink to the source, marked in green color in Fig. \ref{dag_ex2}(c). Each vertex added by zero-padding is marked by red color. The eigenvalues of the obtained zero-padded graph are given in Fig. \ref{dag_ex2_ev} (left). They are all distinct, ensuring the diagonalizability of the adjacency matrix. The relation between the eigenvalue index and the discrete frequency value, calculated as an angle of the corresponding eigenvalue is presented in Fig. \ref{dag_ex2_ev} (right). It can be seen that this relation is close to the linear one. The signal processed on this zero-padded graph with a system of up to the third order would remain the same as in the original DAG, where the Fourier analysis was not possible. 

Finally, we have generated a random uniformly distributed signal from 0 to 1 with 20 samples, $\mathbf{x}=\mathrm{rand}(20,1)$ on the graph given in Fig. \ref{dag_ex2}(a). The graph and the signal are zero-padded according to the presented procedure (see Fig. \ref{dag_ex2}(c)). Then the GFT of the zero-padded signal $\mathbf{x}_{ZP}$ is calculated as $\mathbf{X}_{ZP}=\mathbf{V}^{-1}\mathbf{x}_{ZP}$. The obtained GFT is multiplied by the system transfer function
$$ \mathbf{H}(\mathbf{\Lambda})=h_0\mathbf{I}+ h_1\mathbf{\Lambda}+  h_2\mathbf{\Lambda}^2 +  h_3\mathbf{\Lambda}^3
$$
in order to obtain the GFT of the output signal $ \mathbf{Y}_{ZP}= \mathbf{H}(\mathbf{\Lambda})\mathbf{X}_{ZP}$. The output signal is calculated using the IGFT, as $\mathbf{y}_{ZP}=\mathbf{V}\mathbf{Y}_{ZP}$. Its values on the vertices from the original DAG are the same as the ones obtained by a direct calculation in the vertex domain.

\begin{figure}[tbp]
	\centering
	\includegraphics[width=\linewidth]{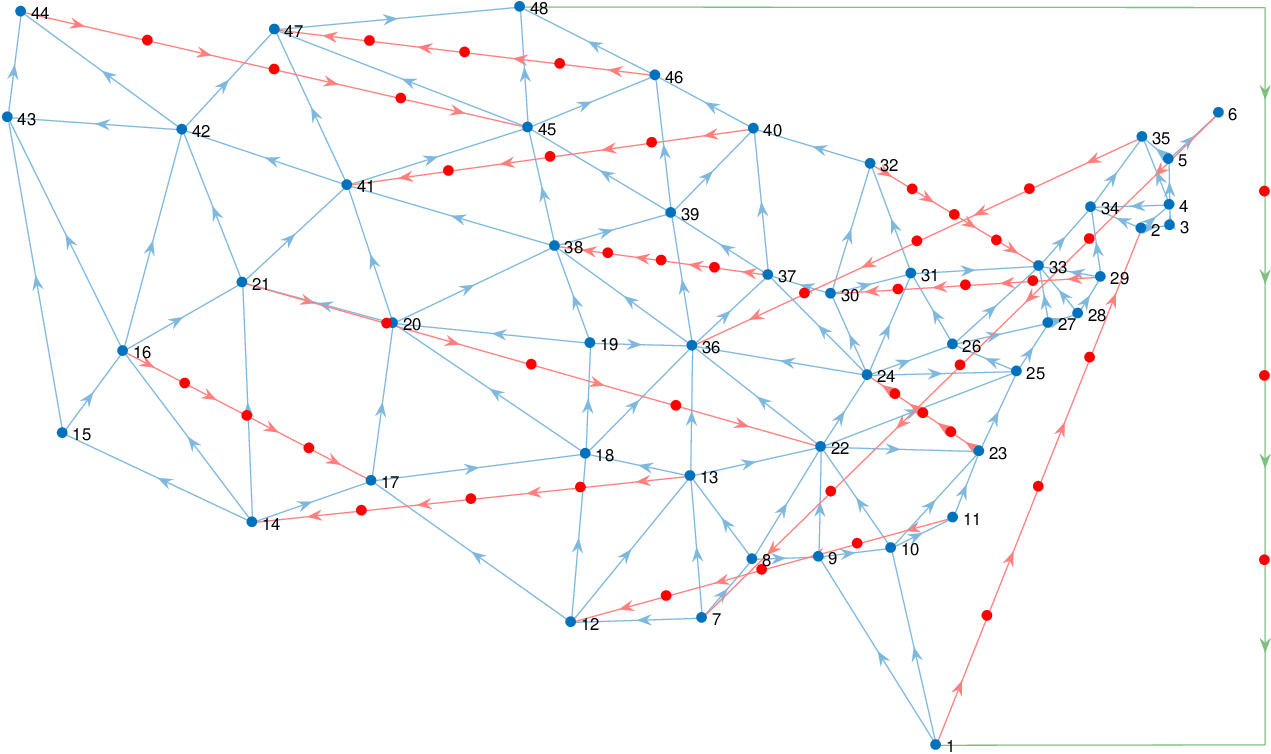}
	\includegraphics[width=\linewidth]{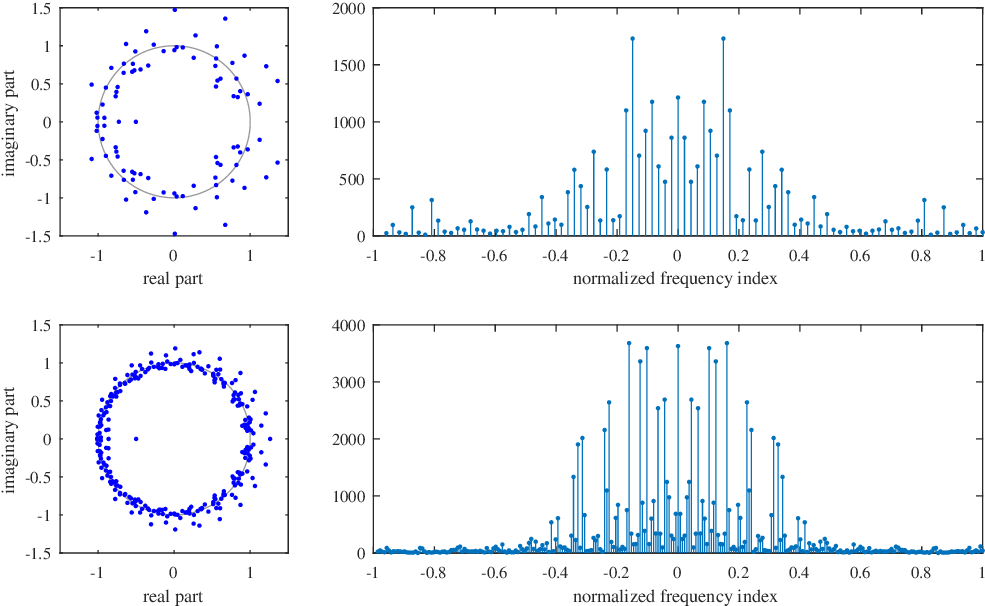}
	\caption{ 
		The USA temperature map \cite{22,27,seifert2021digraph} zero-padded using $M=3$. Edges and vertices added in the zero-padding procedure are marked in red, while the return path is marked in green. Additionally, the eigenvalues (scatter plots  of their imaginary and real part, on the left) and the corresponding GFTs of the temperature signal  (on the right) with $M=3$ (upper) and $M=15$ (lower) are provided. }
	\label{dag_ex5}
\end{figure}

Furthermore, we will consider the USA temperature map, commonly used as a benchmark case for directed graphs \cite{22,27,seifert2021digraph}. The results are shown in Fig. \ref{dag_ex5}. As the number of zero-padded vertices, defined by $M$, increases, it becomes evident that the eigenvalues are located closer to the unit circle (scatter plots on the left). The GFT of the temperature signal calculated using $M=3$ and $M=15$ are also  presented in Fig. \ref{dag_ex5} (right).

\section{Conclusion}

In this paper, we have proposed a  graph zero-padding concept in order to overcome the inherent limitations associated with spectral signal analysis and processing on DAGs. The main idea is motivated by the classical zero-padding technique and its interpretation within the graph signal processing framework.

In the case of a connected DAG, the extension is straightforward. It involves adding a single backward path (or edge), whose length is equal to the maximum order of the system used for graph signal processing. The determinant of the adjacency matrix of the zero-padded graph becomes either $1$ or $-1$, indicating that all eigenvalues are nonzero. For a general form of a DAG, we have presented an algorithm to first ensure the connectivity of the DAG. Following this, we add a backward path to the connected graph.

All changes in the graph structure (including added edges and vertices) are made under the condition that the output signal obtained by a system on a zero-padded graph completely matches the output of the same system on the original DAG. Considering that the number of added vertices can vary, as long as their number is greater than or equal to the order of a system on the graph, we can conclude that diagonalizability can be achieved in practically (almost) all cases of DAGs using the presented approach.

\bibliographystyle{elsarticle-num} 
\bibliography{gzp.bib}

\end{document}